\documentclass[12pt]{article}
\usepackage{graphicx}
\usepackage{epstopdf}
\usepackage{epsfig}
\usepackage{amsmath}
\usepackage{slashbox}
\begin{document}

\title{Perturbations to $\mu-\tau$ symmetry in neutrino mixing}
\author{Jiajun Liao$^1$, D. Marfatia$^2$, and K. Whisnant$^1$\\
\\
\small\it $^1$Department of Physics and Astronomy, Iowa State University, Ames, IA 50011, USA\\
\small\it $^2$Department of Physics and Astronomy, University of Kansas, Lawrence, KS 66045, USA}
\date{}
\maketitle

\begin{abstract}
Many neutrino mixing scenarios that have \(\mu-\tau\) symmetry with $\theta_{13}=0$ are in
disagreement with recent experimental results that indicate a
nonzero value for \(\theta_{13}\). We investigate the effect of small
perturbations on Majorana mass matrices with \(\mu-\tau\)
symmetry and derive analytic formulae for the corrections to the
mixing angles. We find that since \(m_1\) and \(m_2\) are nearly
degenerate, \(\mu-\tau\) symmetry mixing scenarios are able
to explain the experimental data with about the same size
perturbation for most values of $\theta_{12}$. This suggests that the
underlying unperturbed mixing need not have $\theta_{12}$ close to the
experimentally preferred value. One consequence of this is that a new
class of models with \(\mu-\tau\) symmetry is possible, with
unperturbed $\theta_{12}$ equal to zero or $90^\circ$ for arbitrary
unperturbed $\theta_{13}$.
\end{abstract}

\newpage

Of the numerous neutrino mixing scenarios discussed in the
literature~\cite{Hirsch:2012ym}, several have \(\mu-\tau\)
symmetry, 
such as tri-bimaximal mixing (TBM)~\cite{hep-ph/0202074}, bimaximal mixing
(BM)~\cite{hep-ph/9708483}, hexagonal mixing
(HM)~\cite{arXiv:1004.2798} and scenarios of \(A_5\)
mixing~\cite{hep-ph/0306002}. In these scenarios,
\(\theta_{23}=45^\circ\), \(\theta_{13}=0\), and only \(\theta_{12}\)
depends on the particular model. Tri-bimaximal mixing is
most popular because the value of \(\theta_{12}\) predicted
by TBM is close to that preferred by the current experimental
data. However, the latest results from the T2K~\cite{arXiv:1106.2822},
MINOS~\cite{Adamson:2011qu}, and Double Chooz~\cite{Abe:2011fz}
experiments suggest a nonzero value of \(\theta_{13}\), and the recent
Daya Bay~\cite{An:2012eh} and RENO~\cite{Ahn:2012nd} experiments find
\(\theta_{13}\neq 0\) at the \(5.2\sigma\) and \(4.9\sigma\) level,
respectively. Various corrections may reconcile such models with
nonzero $\theta_{13}$~\cite{Hirsch:2012ym}.
In this Letter we consider small perturbations acting on Majorana mass
matrices with \(\mu-\tau\) symmetry and estimate the size of
perturbations required to explain the experimental data.

We find that for \(\mu-\tau\) symmetries with almost any initial value
of $\theta_{12}$ (i.e., before the perturbation), the minimal
size of the perturbations needed to bring the model in agreement with
experimental data varies by only about 20\%. The reason is
that the $\theta_{12}$ correction depends only on the ratio of
perturbation terms and not on their absolute size, and the overall
size of the perturbation is determined by the corrections to
$\theta_{13}$ and $\theta_{23}$, which are relatively small. We also
show that a new category of models with \(\mu-\tau\) symmetry,
\(\theta_{23}=45^\circ\), \(\theta_{12}=0\) or $90^\circ$, and arbitrary \(\theta_{13}\),
can also fit the data with small perturbations.

We start with the mass matrix for Majorana neutrinos
\begin{equation}
M=U^*M^\text{diag} U^\dagger\,,
\end{equation}
where \(M^\text{diag} = \text{diag}(m_1,m_2,m_3)\), \(U\) is the
Pontecorvo-Maki-Nakagawa-Sakata (PMNS) mixing
matrix~\cite{FERMILAB-PUB-10-665-PPD} (without the multiplicative diagonal matrix of Majorana phases), and we work in the basis in which
the charged lepton mass matrix is diagonal. The masses $m_2$ and $m_3$
are complex and $m_1$ can be taken to be real and non-negative.

The general condition describing \(\mu-\tau\) symmetry (also sometimes
called \(\mu-\tau\) universality) is~\cite{Harrison:2002et}
\begin{equation}
|U_{\mu i}|=|U_{\tau i}|\,,\text{ for } i=1,2,3.
\end{equation}
From the standard form of the mixing matrix these conditions are equivalent to
\begin{align}
\theta_{23}=45^\circ\,,\quad
\text{Re}(\cos \theta_{12}\sin\theta_{12}\sin\theta_{13}e^{i\delta})=0\,.
\end{align}
Hence, there are four classes of \(\mu-\tau\) symmetry: (a)
\(\theta_{23}=45^\circ, \theta_{13}=0\); (b) \(\theta_{23}=45^\circ,
\theta_{12}=0\); (c) \(\theta_{23}=45^\circ, \theta_{12}=90^\circ\);
(d) \(\theta_{23}=45^\circ, \delta=\pm 90^\circ\). Class (a) contains
models with tri-bimaximal, bimaximal, hexagonal, and $A_5$ symmetries,
while class (d) includes tetramaximal symmetry~\cite{arXiv:0805.0416}. Classes (b) and
(c) have not been studied before because the unperturbed $\theta_{12}$
angle is far from the experimentally preferred value, but, as we show
below, small perturbations can have a large effect on $\theta_{12}$,
and therefore these models should not be ignored.

\newpage
{\it Class (a): $\theta_{23}^0 = 45^\circ$, $\theta_{13}^0 = 0$}
~\\

We first examine the effect of small perturbations on models in class
(a). The initial (unperturbed) mixing matrix can be written as
\begin{equation}
U_0=\begin{pmatrix}
   \cos\theta_{12}^0 & \sin\theta_{12}^0 & 0 \\[0.3em]
   -\frac{\sin\theta_{12}^0}{\sqrt{2}} & \frac{\cos\theta_{12}^0}{\sqrt{2}} & \frac{1}{\sqrt{2}} \\[0.3em]
   \frac{\sin\theta_{12}^0}{\sqrt{2}} & -\frac{\cos\theta_{12}^0}{\sqrt{2}} & \frac{1}{\sqrt{2}}
   \end{pmatrix}\,,
\end{equation}
and the initial mass matrix is
\begin{align}
&M_0=U_0^*M_0^\text{diag}U_0^\dagger= \nonumber \\
&\begin{pmatrix}
   m_1^0c_{12}^2+m_2^0s_{12}^2 & \frac{(m_2^0-m_1^0)s_{12}c_{12}}{\sqrt{2}} & \frac{(m_1^0-m_2^0)s_{12}c_{12}}{\sqrt{2}} \\
   \frac{(m_2^0-m_1^0)s_{12}c_{12}}{\sqrt{2}} & \frac{1}{2}(m_3^0+m_2^0c_{12}^2+m_1^0s_{12}^2) & \frac{1}{2}(m_3^0-m_2^0c_{12}^2-m_1^0s_{12}^2)  \\
   \frac{(m_1^0-m_2^0)s_{12}c_{12}}{\sqrt{2}} & \frac{1}{2}(m_3^0-m_2^0c_{12}^2-m_1^0s_{12}^2)  & \frac{1}{2}(m_3^0+m_2^0c_{12}^2+m_1^0s_{12}^2)
   \end{pmatrix}\,,
\label{eq:M0a}
\end{align}
where \(M_0^\text{diag}=\text{diag}(m_1^0, m_2^0, m_3^0)\), and
\(c_{jk}\), \(s_{jk}\) denotes \(\cos\theta_{jk}^0\) and
\(\sin\theta_{jk}^0\) respectively. Under a small perturbation
the final (resultant) mass matrix can be written as
\begin{equation}
M=U_0^*M_0^\text{diag}U_0^\dagger+E\,,
\end{equation}
where the perturbation matrix \(E\) has the general form
\begin{equation}
E= M - M_0 = \begin{pmatrix}
   \epsilon_{11} & \epsilon_{12} & \epsilon_{13} \\
   \epsilon_{12} & \epsilon_{22} & \epsilon_{23} \\
   \epsilon_{13} & \epsilon_{23} & \epsilon_{33}
   \end{pmatrix}\,.
\end{equation}

Treating the three masses as eigenvalues of the mass matrix with
each column of the mixing matrix as the corresponding eigenvector,
we can use traditional perturbation methods to find the corrections
to the three angles and three masses. From experiment we know
that \(m_1\) and \(m_2\) are nearly degenerate, so that degenerate
perturbation theory with $|\delta m_{21}^0| \ll |\delta m_{31}^0|$ and
$|\epsilon_{ij}| < |m_k^0|$ (where $\delta m_{ji}^0=m_j^0-m_i^0$, and 
the index $k$ denotes the heaviest eigenstate), can be used.
For simplicity, we assume $M_0$ and $E$
are real and employ the following notation:
\begin{align}
\epsilon_1&=\epsilon_{11}\,,\quad
\epsilon_2=\epsilon_{12}+\epsilon_{13}\,,\quad
\epsilon_3=\epsilon_{12}-\epsilon_{13}\,,\quad
\epsilon_4=\epsilon_{22}+\epsilon_{33}+2\epsilon_{23}\,, \nonumber \\
\epsilon_5&=\epsilon_{22}-\epsilon_{33}\,,\quad
\epsilon_6=\epsilon_{22}+\epsilon_{33}-2\epsilon_{23}-2\epsilon_{11}\,.
\label{eq:epsilons}
\end{align}
We find the first order corrections to the three masses to be
\begin{equation}
\delta m_i^{(1)} = \frac{1}{4}\bigg[4\epsilon_1+\epsilon_6 \pm\bigg(2\delta m_{21}^0
-\sqrt{8\epsilon_3^2 + \epsilon_6^2 + 4(\delta m_{21}^0)^2
+ 4\delta m_{21}^0(2\sqrt{2}\epsilon_3 \sin2\theta_{12}^0
+ \epsilon_6\cos2\theta_{12}^0)}\bigg)\bigg]\,,
\end{equation}
where the plus sign is for $i=1$ and the minus sign is for $i=2$, and
\begin{equation}
\delta m_3^{(1)} = \frac{1}{2}\epsilon_4\,.
\end{equation}
The first order corrections to the mixing angles are
\begin{align}
\delta\theta_{12}^{(1)} &= \frac{1}{2}\arctan
\frac{2\sqrt{2}\epsilon_3 \cos2\theta_{12}^0
-\epsilon_6\sin2\theta_{12}^0}
{2\sqrt{2}\epsilon_3 \sin2\theta_{12}^0
+\epsilon_6\cos2\theta_{12}^0+2\delta m_{21}^0}\,,
\label{eq:dtheta12}
\\
\delta\theta_{23}^{(1)} &=
\frac{\epsilon_5 s_{12}^2 -\sqrt{2}\epsilon_2s_{12}c_{12}}{2\delta m_{31}^0}
+\frac{\epsilon_5 c_{12}^2 +\sqrt{2}\epsilon_2s_{12}c_{12}}{2\delta m_{32}^0}\,,
\label{eq:dtheta23}
\\
\delta\theta_{13}^{(1)} &=
\frac{\sqrt{2}\epsilon_2 c_{12}^2-\epsilon_5s_{12}c_{12}}{2\delta m_{31}^0}
+\frac{\sqrt{2}\epsilon_2 s_{12}^2+\epsilon_5s_{12}c_{12}}{2\delta m_{32}^0}\,,
\label{eq:dtheta13}
\end{align}
and the second order correction to $\theta_{12}$ is
\begin{align}
\delta\theta_{12}^{(2)} =
-\frac{\sqrt{2}\epsilon_2\epsilon_5\cos2(\theta_{12}^0+\delta\theta_{12}^{(1)})
+(\epsilon_2^2-\epsilon_5^2/2) \sin2(\theta_{12}^0 +\delta\theta_{12}^{(1)})}
{4 \delta m_{21}^0 \delta m_{32}^0}\,.
\end{align}
Imposing \(|\delta m_{21}^0| \ll |\delta m_{31}^0|\), the expressions for
\(\delta\theta_{23}^{(1)}\) and \(\delta\theta_{13}^{(1)}\) simplify to
\begin{align}
\delta\theta_{23}^{(1)} \simeq \frac{\epsilon_5}{2\delta m_{31}^0}\,,\quad
\delta\theta_{13}^{(1)} \simeq \frac{\sqrt{2}\epsilon_2}{2\delta m_{31}^0}\,.
\label{eq:approx}
\end{align}
We note that while \(\delta \theta_{23}^{(1)}\) and \(\delta
\theta_{13}^{(1)}\) are suppressed by a factor of order
\(\epsilon_j/\delta m_{31}^0\), to leading order \(\delta
\theta_{12}\) depends only on ratios of linear combinations of
$\epsilon_3$, $\epsilon_6$ and $\delta m_{21}^0$ (which is ${\cal{O}}(\epsilon_{ij})$). Therefore large corrections to
$\theta_{12}$ are possible even for small corrections to $\theta_{23}$
and $\theta_{13}$.

A recent global three-neutrino fit~\cite{arXiv:1205.5254} yields the
parameter values in Table~\ref{tab:data}. We have done a numerical
search to find perturbed mass matrices that give the
oscillation parameters {\it and} which have small
perturbations. In our search, we first fix
\(\theta_{23}^0=45^\circ\) and \(\theta_{13}^0=0\), consistent with
\(\mu-\tau\) symmetry, and choose a particular value for
\(\theta_{12}^0\) and the magnitude of \(m_1\) for the normal
hierarchy (or \(m_3\) for the inverted hierarchy). The global fit in
Table~\ref{tab:data} then defines the magnitudes of the other two
final masses and the three final mixing angles (since
\(\theta_{13}^0=0\), the initial Dirac phase does not matter).

\begin{table}
\caption{Best-fit values and $2\sigma$ ranges of the oscillation parameters~\cite{arXiv:1205.5254} used to find the $\epsilon_{ij}$, with $\delta m^2 \equiv |m_2|^2-m_1^2$ and $\Delta m^2 \equiv |m_3|^2-(m_1^2+|m_2|^2)/2$.
\label{tab:data}}
\begin{center}
\begin{tabular}{|l|*{5}{c|}}\hline
\makebox[4em]{Hierarchy}
&\makebox[2em]{$\theta_{12}(^\circ)$}&\makebox[2em]{$\theta_{13}(^\circ)$}&\makebox[2em]{$\theta_{23}(^\circ)$}
&\makebox[6em]{$\delta m^2(10^{-5}\text{eV}^2)$}&\makebox[7em]{$|\Delta m^2|(10^{-3}\text{eV}^2)$}\\\hline
Normal &$33.6^{+2.1}_{-2.0}$&$8.9^{+0.9}_{-0.9}$&$38.4^{+3.6}_{-2.3}$&$7.54^{+0.46}_{-0.39}$&$2.43^{+0.12}_{-0.16}$\\\hline
Inverted &$33.6^{+2.1}_{-2.0}$&$9.0^{+0.8}_{-1.0}$&$38.8^{+5.3}_{-2.3} \oplus 47.5-53.2$&$7.54^{+0.46}_{-0.39}$&$2.42^{+0.11}_{-0.16}$\\\hline
\end{tabular}
\end{center}
\end{table}


We characterize the size of the perturbation as the
root-mean-square (RMS) value of the perturbations, i.e.,
\begin{equation}
\epsilon_{RMS} = \sqrt{\frac{\sum_{i,j=1}^3 |M_{ij}-M_{0ij}|^2}{9}}\,,
\label{eq:rms}
\end{equation}
where $i$ and $j$ sum over neutrino flavors.  
Hence, $\epsilon_{RMS}$ is determined by the following quantities:
three initial masses, two initial Majorana phases, two final Majorana phases 
and one final Dirac phase.  
We scan over these quantities with all phases taken to be either 0 or 180$^\circ$ to find the 
minimum value of $\epsilon_{RMS}$ for a given $\theta_{12}^0$.
We follow the same procedure for classes (b) and (c) below. For class (d), all
values of the phases are allowed.

We show the perturbations that give the smallest $\epsilon_{RMS}$ for
the normal hierarchy, $m_1 = 0$ and several values of $\theta_{12}^0$
in Table~\ref{tab:normal}.  It is clear that the sizes of $\epsilon_{RMS}$
are approximately the same regardless of the value of
$\theta_{12}^0$; we find that the smallest $\epsilon_{RMS}$ for each
$\theta_{12}^0$ varies by at most 17\% for the examples shown. This
can be explained by the perturbation results derived above as
follows. From Eq.~(\ref{eq:epsilons}) we have
$\epsilon_{RMS}=\sqrt{\epsilon_1^2 + \epsilon_2^2 + \epsilon_3^2
+ \frac{1}{2}\epsilon_5^2 + \frac{1}{4}\epsilon_4^2 + \frac{1}{4}(2\epsilon_1
+\epsilon_6)^2}/3$;
since \(m_3\gg m_1,m_2\) for the normal hierarchy with \(m_1=0\) eV
and the first order perturbations of the three masses are much smaller
than \(m_3\), we can assume \(\delta m_{31}^0\approx m_3^0\approx
m_3\approx\sqrt{ \Delta m^2}=0.0493\) eV. Then from
Eq.~(\ref{eq:approx}) we know that in order to get the correction
\(\delta\theta_{23}=-6.6^\circ\) and \(\delta\theta_{13}=8.9^\circ\)
for any value of $\theta_{12}^0$,
we need \(\epsilon_5=-0.0114\)~eV and
\(\epsilon_2=0.0108\)~eV, so that
\(\sqrt{\epsilon_2^2+\epsilon_5^2/2}/3 = 0.00449\)~eV, which is already
close to the $\epsilon_{RMS}$ values found in
Table~\ref{tab:normal}. The small discrepancy can be explained by the
perturbation of the three masses and other \(\epsilon\)'s.
Hence, we can say that the size of the perturbation mainly comes from
the corrections to \(\theta_{23}\) and \(\theta_{13}\). From
Eq.~(\ref{eq:dtheta12}) we know that the correction to \(\theta_{12}\)
is determined by the relative ratio of \(\epsilon_3\) to
\(\epsilon_6\) and the actual size of the perturbation does not
matter. This means that we can have large corrections for
\(\theta_{12}^0\) with a (relatively) small perturbation.

\begin{table}
\caption{Top half: values of the perturbations (in $10^{-3}$~eV) that
 give the best-fit parameters in Table~\ref{tab:data}
{\it and} have the minimum $\epsilon_{RMS}$ for the given $\theta_{12}^0$,
for the normal hierarchy and $m_1=0$. Bottom half: representative values
that fit the experimental data within $2\sigma$ and for which all
$\epsilon_{ij}$ have a similar magnitude (with $m_1^0 = 0$, $m_2^0 = 0.0054$~eV,
$m_3^0 = 0.0595$~eV, $m_1=0.0072$~eV, $\delta=180^\circ$ and all other phases equal to 0). \label{tab:normal}}
\begin{center}
\begin{tabular}{|l|*{7}{c|}}\hline
$\theta_{12}^0(^\circ)$
&\makebox[3em]{$\epsilon_{11}$}
&\makebox[3em]{$\epsilon_{12}$}
&\makebox[3em]{$\epsilon_{13}$}
&\makebox[3em]{$\epsilon_{22}$}
&\makebox[3em]{$\epsilon_{23}$}
&\makebox[3em]{$\epsilon_{33}$}
&\makebox[3em]{$\epsilon_{RMS}$}\\
\hline
$60$ &-3.05&-3.50&-5.99&-2.72&-1.52&5.77&4.10\\
\hline
$45$ (BM) &-1.32&-4.74&-4.74&-3.58&-0.66&4.90&3.79\\
\hline
$35.3$ (TBM) &0.32&-4.66&-4.82&-4.40&0.16&4.08&3.74\\
\hline
$30$ (HM) &1.07&-4.31&-5.18&-4.78&0.54&3.71&3.79\\
\hline
$0$ &0.00&-1.38&-8.11&-4.24&0.00&4.24&4.36\\
\hline
\hline
$60$ &5.41&-4.17&-4.52&-5.00&-9.94&3.36&6.14\\
\hline
$45$ (BM) &6.76&-4.43&-4.26&-5.67&-9.27&2.69&6.08\\
\hline
$35.3$ (TBM) &7.66&-4.32&-4.37&-6.12&-8.82&2.24&6.08\\
\hline
$30$ (HM) &8.11&-4.17&-4.52&-6.35&-8.59&2.01&6.09\\
\hline
$0$ &9.46&-2.52&-6.17&-7.02&-7.92&1.34&6.28\\
\hline
\end{tabular}
\end{center}
\end{table}

We note that initial values of $\theta_{12}$ on the ``dark side''
($\theta_{12}^0 > 45^\circ$ and $m_1^0 < m_2^0$) can also fit the data
with perturbations that are similar in magnitude to those needed for
tri-bimaximal mixing (see the entry for $\theta_{12}^0 = 60^\circ$ in
Table~\ref{tab:normal}).

In the top half of Table~\ref{tab:normal}, $\epsilon_{11}$ and
$\epsilon_{23}$ are much smaller than the other $\epsilon_{ij}$ for
some values of $\theta_{12}^0$. We have checked that if these values
are set to zero, the experimental constraints can still be satisfied
at the $2\sigma$ level without a large change in the nonzero
parameters. Therefore if some perturbations are exactly zero due to
symmetries, the resulting mass matrix can still fit the experimental
data with small perturbations.

For the inverted hierarchy, some representative sets of
$\epsilon_{ij}$ that give the minimum $\epsilon_{RMS}$ are shown in
Table~\ref{tab:inverted} for $m_3 = 0$. The minimum $\epsilon_{RMS}$
as a function of \(\theta_{12}^0\) varies only by about 1\% in this case,
i.e., the minimum $\epsilon_{RMS}$
varies with \(\theta_{12}^0\) even less for the inverted hierarchy
than for the normal hierarchy.

Clearly, if perturbations are large
enough that tri-bimaximal mixing can explain the experimental data,
then other \(\mu-\tau\) mixing scenarios, such as bimaximal, hexagonal
mixing and \(A_5\) mixing, can also explain the experimental data with
about the same size perturbation. Hence, tri-bimaximal mixing has
no special position among the \(\mu-\tau\) symmetry mixing scenarios
when a perturbation is required to fit the experimental data. Also, it
is possible for all the perturbations to have a similar magnitude and
still give the oscillation parameters within their $2\sigma$ ranges;
see the bottom half of Tables~\ref{tab:normal} and \ref{tab:inverted}.

\begin{table}
\caption{Top half: same as Table~\ref{tab:normal}, except for the inverted
hierarchy and $m_3=0$. Bottom half: same as Table~\ref{tab:normal},
except for the inverted hierarchy and $m_1^0 = 0.05$~eV, $m_2^0 = 0.052$~eV,
$m_3^0 = 0$, $m_3=0.002$~eV, and all phases equal to 0. \label{tab:inverted}}
\begin{center}
\begin{tabular}{|l|*{7}{c|}}\hline
$\theta_{12}^0(^\circ)$
&\makebox[3em]{$\epsilon_{11}$}
&\makebox[3em]{$\epsilon_{12}$}
&\makebox[3em]{$\epsilon_{13}$}
&\makebox[3em]{$\epsilon_{22}$}
&\makebox[3em]{$\epsilon_{23}$}
&\makebox[3em]{$\epsilon_{33}$}
&\makebox[3em]{$\epsilon_{RMS}$}\\
\hline
$60$ &-0.86&-4.94&-5.64&5.57&-0.43&-4.72&4.31\\
\hline
$45$ (BM) &-0.47&-5.29&-5.29&5.38&-0.23&-4.91&4.29\\
\hline
$35.3$ (TBM) &-0.05&-5.30&-5.28&5.17&0.03&-5.12&4.28\\
\hline
$30$ (HM) &0.16&-5.23&-5.36&5.07&0.08&-5.22&4.28\\
\hline
$0$ &0.00&-4.47&-6.12&5.15&0.00&-5.15&4.32\\
\hline
\hline
$60$ &-3.56&-4.89&-5.27&5.95&2.67&-3.92&4.49\\
\hline
$45$ (BM) &-3.06&-4.98&-5.18&5.70&2.92&-4.17&4.47\\
\hline
$35.3$ (TBM) &-2.73&-4.94&-5.22&5.54&3.08&-4.34&4.46\\
\hline
$30$ (HM) &-2.56&-4.89&-5.27&5.45&3.17&-4.42&4.46\\
\hline
$0$ &-2.06&-4.28&-5.88&5.20&3.42&-4.67&4.50\\
\hline
\end{tabular}
\end{center}
\end{table}

We also varied the size of the final masses by changing the value of
\(m_1\) in the normal hierarchy and \(m_3\) in the inverted
hierarchy. We find that the minimum $\epsilon_{RMS}$ decreases as the
size of the final masses increases for both the normal and inverted
hierarchies. For the quasi-degenerate hierarchy (in which the
magnitude of the absolute masses is larger than $\sqrt{\Delta
m^2}$) the size of the perturbation can be very small. This can
be explained by the perturbation equations: since \(\delta m_{31}^0\approx
m_3-m_1\approx\Delta m^2/(m_3+m_1)\) for small perturbations, and
\(\Delta m^2\) is fixed by experimental data, then
\(\delta m_{31}^0\) will decrease if the masses increase, and
similarly for \(\delta m_{32}^0\). Then Eqs.~(\ref{eq:dtheta23}) and
(\ref{eq:dtheta13}) show that in order to get the same corrections for
\(\theta_{13}^0\) and \(\theta_{23}^0\), the size of the perturbation
should also decrease.

\newpage

{\it Classes (b) and (c): \(\theta_{23}^0=45^\circ, \theta_{12}^0=0
\text{ or }90^\circ\)}
~\\

For class (b) (\(\theta_{23}^0=45^\circ\), \(\theta_{12}^0=0\)), since the 
Dirac phase is irrelevant, the initial mixing matrix and mass matrix can be written as
\begin{equation}
U_0=\begin{pmatrix}
   \cos\theta_{13}^0  & 0 & \sin\theta_{13}^0\\[0.3em]
   -\frac{\sin\theta_{13}^0}{\sqrt{2}}  & \frac{1}{\sqrt{2}}& \frac{\cos\theta_{13}^0}{\sqrt{2}} \\[0.3em]
   -\frac{\sin\theta_{13}^0}{\sqrt{2}}  & -\frac{1}{\sqrt{2}}& \frac{\cos\theta_{13}^0}{\sqrt{2}}
   \end{pmatrix}\,,
\end{equation}
and the initial mass matrix is
\begin{align}
&M_0=U_0^*M_0^\text{diag}U_0^\dagger= \nonumber \\
&\begin{pmatrix}
   m_1^0c_{13}^2+m_3^0s_{13}^2
& \frac{(m_3^0-m_1^0)s_{13}c_{13}}{\sqrt{2}}
& \frac{(m_3^0-m_1^0)s_{13}c_{13}}{\sqrt{2}} \\
   \frac{(m_3^0-m_1^0)s_{13}c_{13}}{\sqrt{2}}
& \frac{1}{2}(m_2^0+m_3^0c_{13}^2+m_1^0s_{13}^2)
& \frac{1}{2}(-m_2^0+m_3^0c_{13}^2+m_1^0s_{13}^2)  \\
   \frac{(m_3^0-m_1^0)s_{13}c_{13}}{\sqrt{2}}
& \frac{1}{2}(-m_2^0+m_3^0c_{13}^2+m_1^0s_{13}^2)
& \frac{1}{2}(m_2^0+m_3^0c_{13}^2+m_1^0s_{13}^2)
   \end{pmatrix}\,.
\label{eq:M0b}
\end{align}
If we redefine the phase of the wavefunction \(\psi_3\) to
\(-\psi_3\), or change the initial angle \(\theta_{23}^0\) from
\(45^\circ\) to \(135^\circ\) and switch the indices 2
and 3, then the mass matrix in Eq.~(\ref{eq:M0b}) is exactly the
same as that in Eq.~(\ref{eq:M0a}).

For the above initial mass matrix, 
corrections must shift \(\theta_{12}\) from 0 to
\(33.6^\circ\), and \(\theta_{13}\) from the
initial arbitrary angle to \(9.0^\circ\). We used the same scan
procedure as before and searched for the minimum $\epsilon_{RMS}$ for
various values of $\theta_{13}^0$ (see Table~\ref{tab:normal2}). We
find that for \(\theta_{13}^0 < 20^\circ\),  the
data can be explained with about the same size perturbation as was found for class
(a). For example, when \(\theta_{13}^0=0\) for class (b), the
initial mass matrix is the same as \(\theta_{12}^0=0\) for class (a),
and therefore the minimum $\epsilon_{RMS}$ is also the same. In
particular, when \(\theta_{13}^0\) is close to \(9.0^\circ\) in class
(b), the minimum $\epsilon_{RMS}$ is even smaller than the minimum
value for class (a) because the correction to
\(\theta_{13}\) is smaller in this case. Although the correction to
\(\theta_{12}\) is large, it does not affect the size of the perturbation
too much because its size is mainly determined by the
corrections to \(\theta_{13}\) and \(\theta_{23}\), as noted
before. However, for $\delta \theta_{13}$ greater than about
$20^\circ$, the size of the perturbation required to fit the data
becomes larger since $\theta_{13}$ must change by more than
10$^\circ$.

\begin{table}
\caption{Top half: same as Table~\ref{tab:normal}, except for class (b)
($\theta_{12}^0 = 0$). Bottom half: same as Table~\ref{tab:normal},
except for class (b). \label{tab:normal2}}
\begin{center}
\begin{tabular}{|l|*{7}{c|}}\hline
$\theta_{13}^0(^\circ)$
&\makebox[3em]{$\epsilon_{11}$}
&\makebox[3em]{$\epsilon_{12}$}
&\makebox[3em]{$\epsilon_{13}$}
&\makebox[3em]{$\epsilon_{22}$}
&\makebox[3em]{$\epsilon_{23}$}
&\makebox[3em]{$\epsilon_{33}$}
&\makebox[3em]{$\epsilon_{RMS}$}\\
\hline
$0$ &0.00&-1.38&-8.11&-4.24&0.00&4.24&4.36\\
\hline
$5$ &0.48&1.44&-5.28&-4.48&-0.24&4.00&3.27\\
\hline
$10$ &-0.44&4.21&-2.52&-4.02&0.22&4.46&3.06\\
\hline
$15$ &-2.64&6.59&-0.14&-2.92&1.32&5.56&3.90\\
\hline
$20$ &-5.85&8.30&1.57&-1.32&2.93&7.17&5.24\\
\hline
\hline
$0$ &9.46&-2.52&-6.17&-7.02&-7.92&1.34&6.28\\
\hline
$5$ &9.01&1.13&-2.52&-6.80&-7.69&1.56&5.41\\
\hline
$10$ &7.66&4.67&1.02&-6.12&-7.02&2.24&5.22\\
\hline
$15$ &5.47&8.00&4.35&-5.03&-5.93&3.33&5.80\\
\hline
$20$ &2.50&11.00&7.35&-3.54&-4.44&4.82&6.92\\
\hline
\end{tabular}
\end{center}
\end{table}

For class (c) (\(\theta_{23}^0=45^\circ\),
\(\theta_{12}^0=90^\circ\)), we find that switching \(m_1^0\) with
\(m_2^0\) makes the initial mass matrix the same as the initial mass
matrix of class (b). Since we scan all possible values of
\(m_1^0\) and \(m_2^0\), the minimum $\epsilon_{RMS}$ for a given
$\theta_{13}^0$ for class (c) is the same as for class (b).

~\\
{Class (d): \(\theta_{23}^0=45^\circ, \delta^0=\pm 90^\circ\)}
~\\

If we fix \(\theta_{23}^0=45^\circ, \delta^0=\pm 90^\circ\) and vary
both \(\theta_{12}^0\) and \(\theta_{13}^0\), this category
includes mixing scenarios such as the tetramaximal mixing pattern
\((\text{T}^4\text{M})\)~\cite{arXiv:0805.0416}, and the correlative
mixing pattern with $\delta=\pm 90^\circ$~\cite{Xing:2010pn}. For $\theta_{13}^0 < 20^\circ$ and $\theta_{12}^0 \le
45^\circ$, the smallest $\epsilon_{RMS}$ for the normal hierarchy
(with $m_1=0$) varies from $2.29\times10^{-3}$~eV to
$5.26\times10^{-3}$~eV, where the minimum value occurs at
$\theta_{13}^0 = 9^\circ$ and $\theta_{12}^0 = 32^\circ$, and the
maximum value occurs at $\theta_{13}^0 = 20^\circ$ and $\theta_{12}^0
= 0$. Therefore small perturbations can fit the experimental data for
a wide range of $\theta_{12}^0$ and $\theta_{13}^0$ for class (d).

In summary, we studied small perturbations to Majorana mass matrices
with $\mu-\tau$ symmetry that yield experimentally preferred
oscillation parameters. We find that the size of the perturbations
(which decreases as the neutrino mass scale is increased), is mainly
determined by the corrections to $\theta_{23}$ and $\theta_{13}$, and
that small perturbations can give a very large correction to
$\theta_{12}$ because to first order, the $\theta_{12}$ correction
depends only on the ratio of perturbation terms and not on their
absolute size. Hence, most mixing scenarios with $\mu-\tau$ symmetry
can explain the experimental data with perturbations of similar
magnitude, and tri-bimaximal mixing has no special place among
scenarios with $\mu-\tau$ symmetry. We also find that slightly
perturbed $\mu-\tau$ symmetric models with $\theta_{12} = 0$ or
$90^\circ$ are viable for $\theta_{13} < 20^\circ$.


\vskip 0.1in
{\it Acknowledgments:} KW thanks the University of Kansas for its hospitality during the initial
stages of this work. This research was supported by the
U.S. Department of Energy under Grant Nos.~DE-FG02-01ER41155 and
DE-FG02-04ER41308, and by the NSF under Grant No. PHY-0544278.

\end {document}